%% file: main.tex
\documentclass[twocolumn]{lab}

\usepackage[utf8]{inputenc}
\usepackage[T1]{fontenc}
\usepackage{lmodern}
\usepackage{geometry}
\usepackage{amsmath,amssymb}  %
\usepackage{times}

\usepackage[toc,page,header]{appendix}

\usepackage{cleveref}
\usepackage{subcaption}
\usepackage{booktabs}
\usepackage{graphicx}
\usepackage{xcolor}
\usetikzlibrary{patterns}
\usepackage{float}
\usepackage{placeins}
\usepackage{caption}
\usepackage{bm}

\usepackage{parskip}
\usepackage{array,tabularx}
\usepackage{xurl}

\definecolor{cvprblue}{rgb}{0.21,0.49,0.74}
\definecolor{fallbackgreen}{rgb}{130, 180, 102}
\definecolor{stopred}{rgb}{251, 225, 224}

\input{macro}

\DeclareTextFontCommand{\textbf}{\bfseries}

\renewcommand{\thefootnote}{}

\usepackage{xspace}
\DeclareRobustCommand{\our}{SS-HSDP\xspace}
\DeclareRobustCommand{\circled}[1]{\textcircled{\footnotesize #1}}

\hypersetup{
  pdftitle={\our: Efficient and Flexible Sharded Data Parallelism via Runtime--Update Decoupling},
  pdfauthor={Chendong Xiang, Pengle Zhang, Jinliang Li, Zichen Liang, Jintao Zhang, Fan Bao, Jun Zhu, Jianfei Chen},
}

\title{\our: Efficient and Flexible Sharded Data Parallelism via Runtime--Update Decoupling}

\abstract{%
Sharded data parallelism enables memory-efficient training of large models.
While efficient training kernels favor packed, kernel-specific data layouts, optimizers may require variant data structures as optimization units.
Existing sharded data-parallel frameworks couple these two choices, limiting runtime efficiency and optimization flexibility.
To address this, we present \textbf{\our}, a training framework that separates runtime parameters from update parameters, allowing computation layouts and optimization units to be organized independently.
Through shared data views and hierarchical sharding, \our supports this separation while avoiding redundant data movement.
This design supports a broad range of optimizers, including matrix-based
methods such as Muon, and performance optimizations such as fully zero-copy execution.
Experiments on video and language model workloads with MXFP8 training show that \our achieves higher Model FLOPs Utilization (MFU) than FSDP2, veScale-FSDP, and Megatron-FSDP.
At 128 GPUs, MFU improves by up to 21.7\% over the fastest evaluated baseline, with comparable or lower peak memory usage.%
}

\renewcommand{\maketitle}{%
  \twocolumn[%
    \vskip 3mm
    \mymaketitle
    \vskip 4mm
    \beginabstract
    \abstractlist\par
    \vskip 6mm
  ]%
}

\renewcommand{\affiliationformat}[2][]{{\affiliationfont {\selectfont $^{#1}$#2}}}

\author[*]{Chendong Xiang}
\author[*]{Pengle Zhang}
\author{Jinliang Li}
\author{Zichen Liang}
\author{Jintao Zhang}
\author[\ddagger]{Fan Bao}
\author[\ddagger]{Jun Zhu}
\author[\ddagger]{Jianfei Chen}
\affiliation{Tsinghua University, Shengshu Technology \\ \url{https://www.vidu.com/}}

\begin{document}

\maketitle

\renewcommand{\thefootnote}{}
\footnotetext{*Core contributors, co-first authorship. $\ddagger$Advisors.}
\renewcommand{\thefootnote}{\arabic{footnote}}

\input{src/1-intro.tex}

\begin{figure*}[!t]
  \centering
  {\def\svgwidth{0.95\textwidth}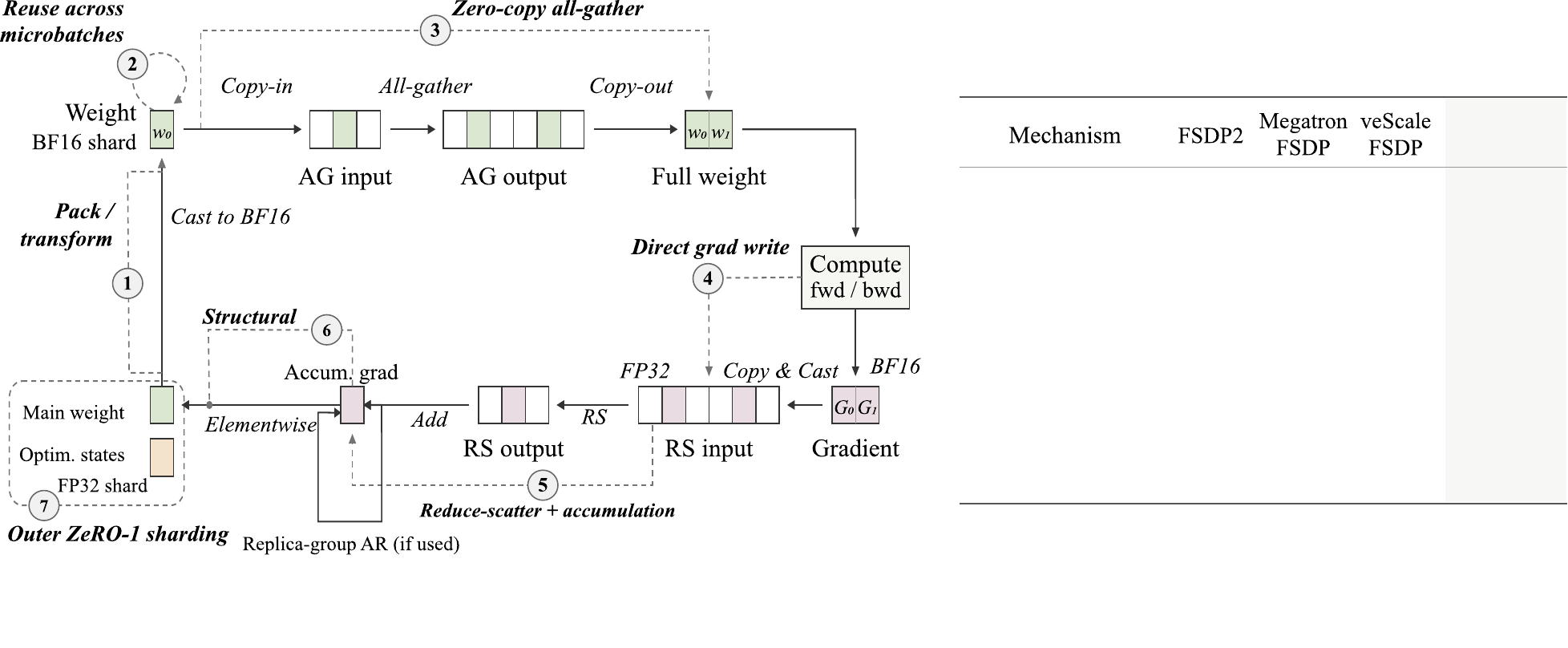}
  \caption{FSDP2-style mixed-precision training with two sharding ranks
  (rank~0 shown). Colored segments track one parameter's weights, gradients,
  and optimizer states; white segments belong to other parameters.
  Solid arrows show the reference dataflow, while numbered dashed paths
  highlight alternative implementations. AG and RS denote all-gather and
  reduce-scatter, respectively.}
  \label{fig:fsdp-dataflow}
\end{figure*}

\input{src/2-background.tex}
\input{src/3-design.tex}
\input{src/4-optimization.tex}
\input{src/5-experiments.tex}

\input{src/6-analysis.tex}

\input{src/7-related-work.tex}
\input{src/8-conclusion.tex}

\bibliographystyle{plain}
\bibliography{references}

\appendix
\input{src/a-implementation.tex}
\input{src/b-evaluation.tex}

\end{document}

%% file: macro.tex
\usepackage{natbib}
\usepackage{latexsym}

\usepackage{url}
\usepackage{amssymb}
\usepackage[utf8]{inputenc}
\usepackage{microtype}
\usepackage{booktabs}
\usepackage{pifont} 
\usepackage{multirow}
\usepackage{makecell}
\usepackage{paralist}
\usepackage{xspace}
\usepackage{color}
\usepackage{xcolor}
\usepackage{colortbl}
\usepackage{adjustbox}
\usepackage{hyperref} 
\usepackage[edges]{forest}
\usepackage{tikz} 
\usepackage{caption}
\usepackage{amsfonts}

\hypersetup{
    colorlinks,
    linkcolor={blue!80!black},
    citecolor={blue!80!black},
}
\tikzset{
    root/.style =             {align=center, text width=1cm, rounded corners=3pt, line width=0.3mm, fill=gray!10, draw=gray!80, font=\small},
    demographic/.style =         {align=center, text width=1.8cm, rounded corners=3pt, line width=0.3mm, fill=blue!10, draw=blue!80, font=\footnotesize},
    demographic_work/.style =    {align=center, text width=10cm, rounded corners=3pt, line width=0.3mm, fill=blue!10, draw=blue!0, font=\footnotesize},
    character/.style =         {align=center, text width=1.8cm, rounded corners=3pt, line width=0.3mm, fill=red!10, draw=red!80, font=\footnotesize},
    character_work/.style =    {align=center, text width=10cm, rounded corners=3pt, line width=0.3mm, fill=red!10, draw=red!0, font=\footnotesize},
    personalization/.style =           {align=center, text width=1.8cm, rounded corners=3pt, line width=0.3mm, fill=cyan!10, draw=cyan!80, font=\footnotesize},
    personalization_work/.style =      {align=center, text width=10cm, rounded corners=3pt, line width=0.3mm, fill=cyan!10, draw=cyan!0, font=\footnotesize},
    risk/.style =         {align=center, text width=1.8cm, rounded corners=3pt, line width=0.3mm, fill=orange!10, draw=orange!80, font=\footnotesize},
    risk_work/.style =    {align=center, text width=10cm, rounded corners=3pt, line width=0.3mm, fill=orange!10, draw=orange!0, font=\footnotesize},
}

\usepackage{CJK}

%% file: src/1-intro.tex
\section{Introduction}
\label{sec:intro}

ZeRO~\cite{zero} enables training large models on memory-limited devices
by partitioning weights, gradients, and optimizer states across
data-parallel ranks. 
Under ZeRO-3, weights are typically all-gathered before each layer's
computation, and gradients are reduce-scattered after its backward
computation. These collectives can overlap with computation in other
layers, potentially hiding most of the communication cost in
compute-bound workloads.
With relatively modest changes to model code,
frameworks such as DeepSpeed~\cite{deepspeed} and
PyTorch FSDP2~\cite{fsdp,fsdp2} make this approach practical for large-scale
training. This approach has been successfully used to train both language models~\cite{llama3,olmo} and video generation models~\cite{wan,cosmos}.

\begin{figure}[!t]
  \centering
  {\def\svgwidth{\columnwidth}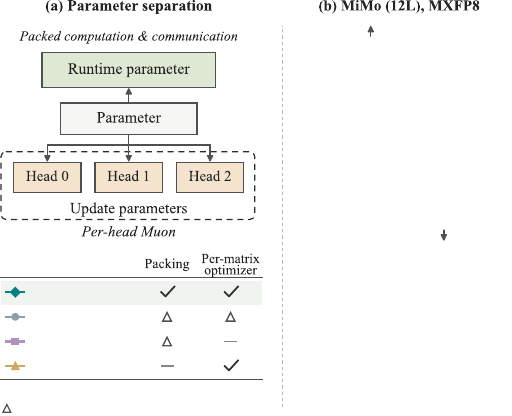}
  \caption{Parameter separation in \our, and MiMo MXFP8
  training performance with AdamW and framework capabilities.}
  \label{fig:intro-decoupling}
\end{figure}

Recent advances in model architectures, low-precision training
and optimization impose new requirements on both runtime computation
efficiency and optimizer flexibility. 
For MoE models~\cite{switch-transformers,deepseek-moe}, redundant movement
of their large weights and gradients can introduce substantial overhead.
Low-precision weight formats such as MXFP~\cite{mxfp} represent weights using quantized values and scale metadata, whose layout must match 
the requirements of high-performance kernels~\cite{deepgemm,cublas-mxfp8}. 
Meanwhile, matrix-based optimizers such as Muon~\cite{muon} operate 
on matrices that may be submatrices of the weights used for computation.
For example, when applying Muon per attention head~\cite{kimi-k3}, each 
head is updated independently, whereas forward and backward computation can use 
one packed QKV projection, with its values and scales arranged for the target GEMM kernel.

These requirements expose a mismatch between how parameters are
computed and how they are updated.
Runtime computation requires parameter data to be arranged for
optimized kernels, packed across multiple parameters and stored in
kernel-specific layouts, whereas optimizers require the same data
to be accessed in the units they optimize, such as individual attention heads.
Tying update units to the parameter organization used for computation
constrains these choices: supporting different update units may require
reorganizing data or customizing the optimizer.

Existing frameworks address only part of this mismatch.
Megatron-FSDP~\cite{megatron-fsdp} optimizes the runtime data path
with flattened buffers, persistent low-precision weights, and direct
gradient writes, but does not support structured optimizer updates.
veScale-FSDP~\cite{vescale-fsdp} aligns ragged shards with the
update and quantization structure to support distributed Muon updates and block-scaled quantization, 
but does not pack or transform parameters into kernel-specific layouts.
Our goal is to satisfy both requirements without trading one for the other.

We present \our,
a framework that decouples \emph{runtime parameters} for efficient computation
from \emph{update parameters} for flexible optimization.
Figure~\ref{fig:intro-decoupling} illustrates this separation and its
performance benefits.
Each update parameter defines a unit of optimization and its
associated operations, including optimizer updates and weight
casting or quantization.
The weights of one or more update parameters
are packed or transformed 
into a runtime parameter's weight in the representation
required by the computation kernel.
With careful memory management and 
hierarchical sharding design,
this separation needs no extra memory operations.
Rather, it preserves update flexibility while
enabling computation- and communication-friendly data layouts
that minimize redundant copies and exploit optimized
kernels and communication operations.

Building on this design, \our further reduces memory allocations and
copies. On supported topologies, it transfers data using copy engines
within nodes and host-driven remote memory access (RMA) between nodes,
reducing contention for streaming multiprocessors (SMs).
As a result, \our achieves up to 1.33$\times$ speedup and 27.3\% memory
saving over PyTorch FSDP2 on 128 GPUs.
In summary, our contributions are:
\begin{itemize}
  \item \textbf{Identify} We identify a mismatch between kernel-specific
        runtime layouts and optimizer update units, and show how coupling
        them limits runtime efficiency and optimizer flexibility.
  \item \textbf{Design} We design a unified framework that separates
        runtime parameters from update parameters, with memory management
        and hierarchical sharding that enable efficient computation and
        flexible optimizer updates without redundant data movement.
  \item \textbf{Implement} A clean, high-performance implementation that supports diverse
        training scenarios and achieves higher MFU than FSDP2,
        veScale-FSDP, and Megatron-FSDP.
\end{itemize}

We will release the code in the future.

%% file: figures/intro-decoupling.pdf_tex
\begingroup%
  \makeatletter%
  \providecommand\color[2][]{%
    \errmessage{(Inkscape) Color is used for the text in Inkscape, but the package 'color.sty' is not loaded}%
    \renewcommand\color[2][]{}%
  }%
  \providecommand\transparent[1]{%
    \errmessage{(Inkscape) Transparency is used (non-zero) for the text in Inkscape, but the package 'transparent.sty' is not loaded}%
    \renewcommand\transparent[1]{}%
  }%
  \providecommand\rotatebox[2]{#2}%
  \newcommand*\fsize{\dimexpr\f@size pt\relax}%
  \newcommand*\lineheight[1]{\fontsize{\fsize}{#1\fsize}\selectfont}%
  \ifx\svgwidth\undefined%
    \setlength{\unitlength}{245.7480011bp}%
    \ifx\svgscale\undefined%
      \relax%
    \else%
      \setlength{\unitlength}{\unitlength * \real{\svgscale}}%
    \fi%
  \else%
    \setlength{\unitlength}{\svgwidth}%
  \fi%
  \global\let\svgwidth\undefined%
  \global\let\svgscale\undefined%
  \makeatother%
  \begin{picture}(1,0.8110764)%
    \lineheight{1}%
    \setlength\tabcolsep{0pt}%
    \put(0,0){\includegraphics[width=\unitlength,page=1]{figures/intro-decoupling.pdf}}%
    \put(0.06069632,0.23379268){\fontfamily{ptm}\fontsize{0.0284844626\unitlength}{0.0356055783\unitlength}\selectfont\color[rgb]{0.14117647,0.14117647,0.14117647}\makebox(0,0)[lt]{\lineheight{1.25}\smash{\begin{tabular}[t]{l}\textbf{\our}\end{tabular}}}}%
    \put(0,0){\includegraphics[width=\unitlength,page=2]{figures/intro-decoupling.pdf}}%
  \end{picture}%
\endgroup%

%% file: figures/fsdp-dataflow-with-table.pdf_tex
\begingroup%
  \makeatletter%
  \providecommand\color[2][]{%
    \errmessage{(Inkscape) Color is used for the text in Inkscape, but the package 'color.sty' is not loaded}%
    \renewcommand\color[2][]{}%
  }%
  \providecommand\transparent[1]{%
    \errmessage{(Inkscape) Transparency is used (non-zero) for the text in Inkscape, but the package 'transparent.sty' is not loaded}%
    \renewcommand\transparent[1]{}%
  }%
  \providecommand\rotatebox[2]{#2}%
  \newcommand*\fsize{\dimexpr\f@size pt\relax}%
  \newcommand*\lineheight[1]{\fontsize{\fsize}{#1\fsize}\selectfont}%
  \ifx\svgwidth\undefined%
    \setlength{\unitlength}{938.6031189bp}%
    \ifx\svgscale\undefined%
      \relax%
    \else%
      \setlength{\unitlength}{\unitlength * \real{\svgscale}}%
    \fi%
  \else%
    \setlength{\unitlength}{\svgwidth}%
  \fi%
  \global\let\svgwidth\undefined%
  \global\let\svgscale\undefined%
  \makeatother%
  \begin{picture}(1,0.41275495)%
    \lineheight{1}%
    \setlength\tabcolsep{0pt}%
    \put(0,0){\includegraphics[width=\unitlength,page=1]{figures/fsdp-dataflow-with-table.pdf}}%
    \put(0.96016373,0.321585){\fontfamily{ptm}\fontsize{0.0145244300\unitlength}{0.0181555375\unitlength}\selectfont\color[rgb]{0,0,0}\makebox(0,0)[t]{\lineheight{1.25}\smash{\begin{tabular}[t]{c}\textbf{\our}\end{tabular}}}}%
    \put(0,0){\includegraphics[width=\unitlength,page=2]{figures/fsdp-dataflow-with-table.pdf}}%
  \end{picture}%
\endgroup%

%% file: src/2-background.tex
\section{Background and Motivation}
\label{sec:background}

\subsection{Notation and ZeRO Training}
\label{sec:notation}

\paragraph{Tensors and shards}
A \emph{tensor} is a multidimensional array of numerical values;
a \emph{matrix} is a two-dimensional tensor.
We use lowercase letters for tensors and corresponding uppercase letters
for \emph{tensor lists}, e.g., $W=[w_0,\ldots,w_k]$.
A tensor list is a collection of tensors with custom layouts.
A \emph{logical tensor} denotes the complete tensor, regardless of
whether all, part, or none of its data are stored locally.

Sharding partitions a tensor or tensor list across a group of ranks,
each holding a local \emph{shard}.
The number of ranks in this \emph{sharding group} is its
\emph{sharding degree}.
When shards are replicated, ranks holding replicas of the same shard
form a \emph{replica group}.
Superscripts identify the sharding level of a local shard,
as defined in the ZeRO sharding paragraph below;
the same notation applies to tensors and tensor lists.
Table~\ref{tab:notation} summarizes the notation.

\paragraph{Parameters and updates}
Following Megatron's naming convention~\cite{megatron-lm,megatron-repo},
a parameter $P$ has several associated tensors for computation and optimization.
In mixed-precision training, a parameter has a \emph{weight} $w$
stored in a low-precision format such as BF16 or FP8, and an FP32 \emph{main weight}
$w^{\mathrm{main}}$ for optimization~\cite{mixed-precision-training}.
Gradients produced by backward computation are reduced across
data-parallel ranks and accumulated over microbatches into a
\emph{main gradient} $g$.
The optimizer uses $g$ to update $w^{\mathrm{main}}$ and its
\emph{optimizer states}, such as momentum $m$.
We consider FP32 main gradients and optimizer states unless otherwise stated.
We denote the optimizer update by $\mathcal{F}$ and weight preparation
by $\mathcal{G}$:
\[
\begin{aligned}
(w^{\mathrm{main}},m) &\xleftarrow{\mathcal{F}}
                         (w^{\mathrm{main}},g,m),\\
(w,s) &= \mathcal{G}(w^{\mathrm{main}}).
\end{aligned}
\]
Depending on the optimizer, additional or different state tensors may be used.
$\mathcal{G}$ casts or quantizes the main weight,
producing optional scale metadata $s$ for quantized formats.

\paragraph{ZeRO sharding}
ZeRO stage 1 shards main weights and optimizer states; stage 2 additionally
shards main gradients; stage 3 further shards computation
weights~\cite{zero}.
We denote local shards at the weight, gradient-accumulation, and
optimizer-update levels by superscripts $\mathrm{z3}$, $\mathrm{z2}$,
and $\mathrm{z1}$, with sharding degrees $s_w$, $s_g$, and $s_o$,
respectively. For example, $w^{\mathrm{z3}}$ and $W^{\mathrm{z3}}$
denote local shards of a weight tensor and a weight tensor list.
Conventional ZeRO-3 uses the same sharding group for all three levels.
Hybrid sharded data parallelism (HSDP) uses smaller sharding groups
and replicates shards across groups to limit repeated weight all-gathers.
Outer ZeRO-1 further shards main weights and optimizer states across
replica groups~\cite{megatron-fsdp}, reducing persistent memory without
enlarging the groups for repeated all-gathers.
With gradients accumulated over microbatches, outer communication
is needed once per optimizer step.

\begin{figure}[!ht]
  \centering
  \includegraphics[width=0.9\columnwidth]{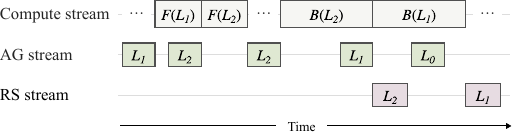}
  \caption{ZeRO-3 Layer-wise($L$) overlap. $F$ and $B$ denote forward
  and backward; AG and RS denote all-gather and reduce-scatter.}
  \label{fig:zero-overlap}
\end{figure}

\paragraph{Training execution}
We call forward and backward computation and their associated communication
\emph{runtime}. Under ZeRO-3, each layer all-gathers weights before
computation and reduce-scatters gradients after backward.
These collectives can overlap with computation in other layers.
Figure~\ref{fig:zero-overlap} illustrates this layer-wise overlap.
After gradients from all microbatches have been accumulated, the
\emph{update} phase applies $\mathcal{F}$ to parameters across the model
and prepares weights through $\mathcal{G}$ for the next training step.

\begin{table}[t]
  \caption{Notation.}
  \label{tab:notation}
  \centering\small
  \setlength{\tabcolsep}{5pt}
  \begin{tabular}{@{}p{0.30\columnwidth}p{0.65\columnwidth}@{}}
    \toprule
    Symbol & Meaning \\
    \midrule
    $P$, $P'$ & Update and runtime parameters \\
    $i$ & Index of an update parameter inside a runtime parameter \\
    $w$, $s$ & Weight and quantization scales \\
    $w^{\mathrm{main}}$, $g$ & Main weight and main gradient \\
    $m$ & Optimizer state, e.g., momentum \\
    $u$, $v$ & Update temporaries \\
    $w'$, $s'$, $g'$ & Runtime weight, scales, and gradient \\
    $W,S,G,M,U,V$ & Lists of corresponding tensors \\
    $W',S',G'$ & Lists of corresponding runtime tensors \\
    \midrule
    $\mathcal{F}$ & Optimizer update \\
    $f_j$ & The $j$th operation in $\mathcal{F}$ \\
    $\mathcal{G}$ & Weight casting or quantization \\
    $\mathcal{T}_w$ &  Weight transform $(w_i,s_i)_i \rightarrow (w',s')$\\
    $\mathcal{T}_g$ &  Gradient transform $g' \rightarrow (g_i)_i$\\
    \midrule
    $\mathrm{z3},\mathrm{z2},\mathrm{z1}$ & ZeRO-3/2/1 shard superscripts \\
    $s_w,s_g,s_o$ & Sharding degrees at $\mathrm{z3},\mathrm{z2},\mathrm{z1}$ \\
    \bottomrule
  \end{tabular}
\end{table}

\subsection{FSDP2 Dataflow}
\label{sec:fsdp2-dataflow}
Figure~\ref{fig:fsdp-dataflow} shows 
PyTorch FSDP2's BF16 mixed-precision
dataflow with FP32 gradient reduction and accumulation~\cite{fsdp2}.
FSDP2 shards each parameter along dimension 0 by default and represents
its shards as DTensors~\cite{dtensor}.
Hybrid sharded data parallelism (HSDP) additionally replicates these
shards in replica groups.
Before forward and backward computation, $\mathcal{G}$ casts
$(w^{\mathrm{main}})^{\mathrm{z3}}$ to BF16 $w^{\mathrm{z3}}$.
Copy-in packs these shards into the all-gather input $W^{\mathrm{z3}}$.
All-gather produces $W$ in an interleaved layout: data are ordered by
rank, leaving each parameter's shards separated by other parameters' data.
Copy-out assembles each contiguous weight $w$ for computation.
Backward produces per-parameter gradients, which are converted to FP32
and packed by destination rank into a reduce-scatter buffer.
HSDP additionally all-reduces the results across replica groups.
The reduced shards initialize $g^{\mathrm{z2}}$ on the first microbatch
and are added to it thereafter.
After all microbatches, $\mathcal{F}$ updates the local main-weight shards $(w^{\mathrm{main}})^{\mathrm{z3}}$.

As indicated by the circled numbers in Figure~\ref{fig:fsdp-dataflow},
FSDP2's default communication path incurs repeated weight
casts~\circled{2}, buffer copies~\circled{3}\circled{4}, and separate
gradient accumulation~\circled{5}.
Its parameter-wise sharding also requires additional handling for
structured optimizer updates~\circled{6} and packed or transformed
kernel-specific layouts~\circled{1}, while HSDP replicates main weights
and optimizer states rather than further sharding them~\circled{7}.
We examine these limitations below.

\subsection{Challenges and Analysis}
\label{sec:challenges}
In conventional BF16 training with AdamW, $w$, $w^{\mathrm{main}}$,
$g$, and the optimizer states have matching shapes, while both
$\mathcal{F}$ and $\mathcal{G}$ operate elementwise.
These tensors can therefore be sharded identically at arbitrary element
boundaries, without considering matrix or block structure.
Structured updates and block quantization change these requirements.
Muon is a matrix-based optimizer~\cite{muon}; applying it per attention
head requires updating each head matrix independently~\cite{kimi-k3},
rather than treating a packed projection as one update unit~\circled{6}.
Similarly, MXFP quantization in $\mathcal{G}$ computes a shared scale
for a block of weight values~\cite{mxfp}.
These operations require complete matrices, heads, or quantization blocks;
when sharded, metadata must identify the fragments needed for communication.

Runtime computation instead favors packing Q, K, and V projections into one GEMM,
with weights and scales arranged in kernel-specific
layouts~\cite{deepgemm,cublas-mxfp8}~\circled{1}.
Preparing these representations for every microbatch can be costly.
Reusing the prepared weight shards avoids this repeated work~\circled{2},
but increases persistent memory usage.
This additional memory cost can be offset by outer ZeRO-1~\circled{7}.
For communication, compatible contiguous buffer layouts let collectives
and computation kernels share storage, avoiding weight and gradient
copies~\circled{3}\circled{4}.
However, evenly sharding these buffers can split matrices, heads,
or quantization blocks required by updates.
The challenge here is to preserve runtime efficiency without restricting
update flexibility or introducing redundant data movement.

Existing frameworks address these requirements in different ways.
Megatron-FSDP~\cite{megatron-fsdp} uses Transformer Engine for
kernel-specific low-precision weight transformations~\circled{1}
and retains the prepared weight shards across microbatches~\circled{2}.
Its flattened buffers avoid weight layout copies~\circled{3}, while
compatible kernels write gradients directly into these buffers~\circled{4}.
It also supports outer ZeRO-1 sharding~\circled{7}, but does not support
structured optimizer updates~\circled{6}.
veScale-FSDP~\cite{vescale-fsdp} uses structure-aware ragged sharding to
accommodate quantization blocks and distributed Muon updates~\circled{6},
and DBuffer to avoid weight layout copies~\circled{3}.
However, neither framework decouples update units from runtime layouts,
limiting the combination of update flexibility and runtime efficiency.

\our decouples update units from runtime layouts, supporting structured
optimizer updates~\circled{6} while packing and transforming weights
for efficient computation~\circled{1}.
Shared buffers and direct gradient writes avoid redundant weight and
gradient copies~\circled{3}\circled{4}.
Prepared weight shards are reused across microbatches to avoid repeated
conversion~\circled{2}, while further sharding main weights and optimizer
states across replica groups can offset the additional memory cost~\circled{7}.
Custom ZeroSM reduce-scatter accumulates directly
into $g$ without a separate post-reduction add~\circled{5}.
Section~\ref{sec:design} presents the parameter separation, memory management,
and sharding design behind these optimizations.

%% file: src/3-design.tex
\section{Design}
\label{sec:design}

\begin{figure*}[!t]
  \centering
  \includegraphics[width=\textwidth]{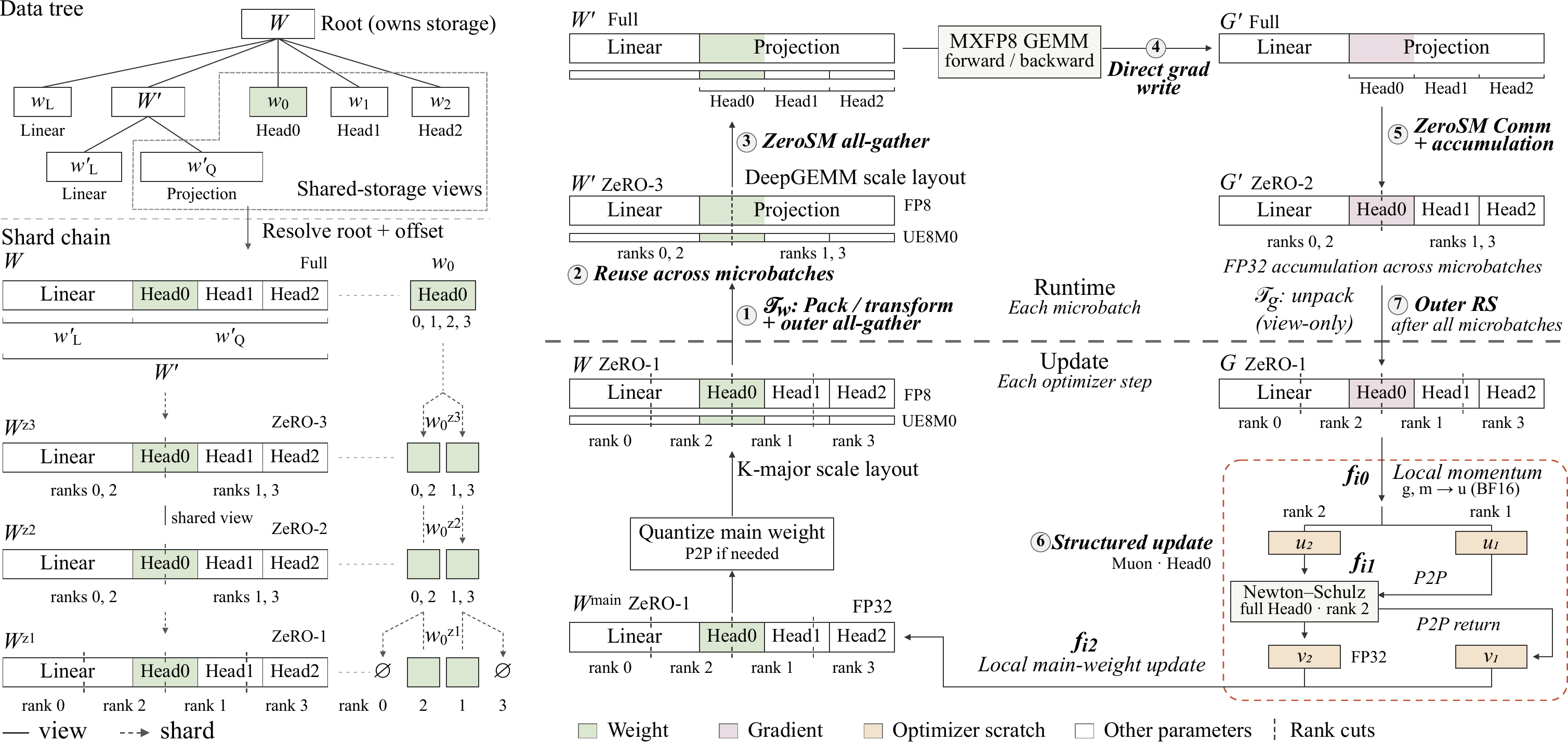}
  \caption{\our data structures and execution flow. Left: data-tree views
  and the Head0 shard chain. Right: packed runtime computation and structured
  updates. The example uses four ranks, with $s_o = 4, s_g = s_w = 2$.}
  \label{fig:fhsdp-design}
\end{figure*}

In this section, we present the core design of \our,
including the separation of update and runtime parameters,
data trees and shard chains for managing flattened buffers and
hierarchical sharding, and parameter grouping.

\subsection{Update and Runtime Parameters}
To support runtime efficiency and update flexibility, \our
divides the parameter into an \emph{update parameter} and a \emph{runtime parameter}, denoted by $P$ and $P'$, respectively.
An update parameter defines an independent unit of optimization, with
the tensors and weight preparation operation $\mathcal{G}$ defined in
Section~\ref{sec:notation}.
We express its optimizer update as an ordered sequence
$\mathcal{F}=[f_0,f_1,\ldots,f_n]$, with optional temporary tensors
such as $u$ and $v$. Each operation specifies the tensors it reads and writes.
Although it contains the 
low-precision representation $(w,s)$, an update
parameter is not directly used by forward or backward kernels.

A runtime parameter $P'$ contains one or more update parameters $P_0,\ldots,P_{k}$ 
whose weight, scale, and main-gradient tensors are denoted by $w_i$, $s_i$, and $g_i$, respectively.
$P'$ holds the runtime weight $w'$, optional scales $s'$, and the gradient $g'$
produced by backward computation.
The transform operations $\mathcal{T}_w$ and $\mathcal{T}_g$
connect the two representations:
\[
\begin{aligned}
(w',s') &= \mathcal{T}_w(w_0,s_0,\ldots,w_k,s_k),\\
(g_0,\ldots,g_k) &= \mathcal{T}_g(g').
\end{aligned}
\]
$\mathcal{T}_w$ prepares the layout required by computation kernels,
while $\mathcal{T}_g$ maps the runtime
gradient back to the main gradients for optimizer updates.

\paragraph{EG1} We use MXFP8 Q projection with per-head Muon updates as an example.
Each Q head has an update parameter $P_i$, while the Q
projection is represented by one runtime parameter $P'$.
For each head, $\mathcal{F}_i = [f_{i0}, f_{i1}, f_{i2}]$, 
$f_{i0}$ updates $m_i$ using $g_i$ and writes a BF16 intermediate to $u_i$.
$f_{i1}$ applies Newton--Schulz to $u_i$ and writes the FP32 update
direction to a separate scratch tensor $v_i$. Then $f_{i2}$ updates
$w_i^{\mathrm{main}}$ using $v_i$.
$\mathcal{G}_i$ uses $32 \times 32$ block quantization in a k-major
layout to produce FP8 $w_i$ and UE8M0 $s_i$.
Then $\mathcal{T}_w$ packs the head weights $w_i$ and relayout
the scales $s_i$ into the layout required by DeepGEMM's MXFP8
kernels for the whole Q projection.
This avoids separate per-head GEMM launches.
$\mathcal{T}_g$ maps $g'$ back to the
per-head $g_i$.

Packing and unpacking may introduce redundant copies and allocations
when the tensors could instead share storage.
The update operations must also remain valid when these buffers
are sharded. We use data trees and shard chains to address
these requirements.

\subsection{Data Tree and Shard Chain}
\our stores each tensor list in a contiguous buffer, grouping tensors
with the same device and dtype for memory management and communication.

\paragraph{Data Tree}
A data tree allows update and runtime tensors to share storage
despite having different shapes and boundaries.
Each tensor or tensor list is represented by a \emph{data node}.
A parent--child link defines a view at a byte offset within the parent;
the child's shape and dtype determine the view's size.
Only the root owns the allocation, and each child accesses this storage
through its accumulated offset.

In EG1, the head weights $w_i$ are consecutive views of $W$,
while the runtime weight $w'$ views their combined region.
The heads can therefore be updated independently and used together
as one projection without moving their weight data.
The weight-packing part of $\mathcal{T}_w$ requires neither a copy
nor an additional allocation, since the packed weight is merely a view
of the root allocation, as shown on the left of Figure~\ref{fig:fhsdp-design};
scale relayout for $s'$ is still
performed to satisfy the kernel's layout requirements.

\paragraph{Shard Chain}
\our partitions each flattened tensor list into contiguous chunks
and further partitions each local chunk at subsequent levels,
recording these partitions as a \emph{shard chain}:
\[
W \longrightarrow W^{\mathrm{z3}}
  \longrightarrow W^{\mathrm{z2}}
  \longrightarrow W^{\mathrm{z1}}.
\]
When adjacent levels use the same sharding degree, their local shards
cover the same data and can share storage through the data tree.
A shard-chain link alone does not create a parent--child link in the data tree.

Each member tensor inherits this chain: its local fragment is the
intersection of its range with the local buffer range, or empty if
they do not overlap.
\[
w_i \longrightarrow w_i^{\mathrm{z3}}
    \longrightarrow w_i^{\mathrm{z2}}
    \longrightarrow w_i^{\mathrm{z1}}.
\]
Each link records the sharding group and partition, mapping local
fragments to their positions in the original tensor.
The same construction applies to all tensor fields.
Together with rank ownership, these mappings locate the fragments
required for distributed updates.

The shard chain can also include other partitions. This section focuses
on hierarchical data-parallel sharding rather than its composition with
tensor or expert parallelism.

\paragraph{Distributed updates}
For each operation in $\mathcal{F}$ and for $\mathcal{G}$, we declare
its input and output tensors and an \emph{update unit size} for each
operand: the number of elements required together, from one element
for elementwise operations to a complete matrix for matrix operations.
\our uses the shard mappings to locate the corresponding
operand fragments at the ZeRO-1 level.
When all required operands for a unit are local, the operation runs
on local views. Otherwise, an execution rank receives the required
fragments through point-to-point (P2P) communication, performs the operation, and returns
the modified fragments to their owners. 
Only declared tensors participate in this exchange.

In EG1, $f_{i0}$ and $f_{i2}$ run locally; only $f_{i1}$ requires the
complete head intermediate $u_i$. Main weights and momentum remain
sharded, and shard boundaries need not coincide with head boundaries.
The data tree manages allocations and shared views, while the shard
chain preserves each tensor's logical positions across sharding levels.
Together, they preserve packed, evenly sharded runtime
buffers without restricting optimizer update units.

\subsection{Parameter Groups}
\our groups compatible parameters for shared storage and batched execution.
Parameters with different dtypes, quantization formats, or optimizers
may require separate tensor lists.
The group boundaries for storage, runtime, and updates need not coincide.

An \emph{atom group} contains compatible runtime parameters and their
update parameters. It packs matching tensor fields into separate tensor
lists and constructs their data trees and shard chains.
A \emph{runtime group} schedules operations across one or more atom
groups, handling coalesced communication, gradient accumulation at
ZeRO-2, and the transforms $\mathcal{T}_w$ and $\mathcal{T}_g$.
These transforms can be configured at different sharding levels.
An \emph{update group} schedules optimizer updates $\mathcal{F}$ and
weight preparation $\mathcal{G}$ at ZeRO-1, batching compatible
operations across its atom groups.
This organization supports layer-wise runtime execution and model-wide
batching of compatible updates.

\subsection{Execution Flow}
We now follow EG1 through the workflow in Figure~\ref{fig:fhsdp-design}.
The example uses four ranks, with ZeRO-1 degree four and ZeRO-2 and
ZeRO-3 degree two. We place $\mathcal{T}_w$ at ZeRO-3 and
$\mathcal{T}_g$ at ZeRO-2.
During setup, \our builds the groups, data trees, and shard chains,
and initializes or loads the main weights and optimizer states.
Before the first training step, $\mathcal{G}$ initialize
weights and scales at ZeRO-1.

\paragraph{Runtime}
At the first forward pass after initialization or an optimizer update,
the runtime group performs the outer all-gather on the FP8 weights
and UE8M0 scales prepared by $\mathcal{G}$ at ZeRO-1.
At ZeRO-3, $\mathcal{T}_w$ prepares packed runtime weights and
kernel-specific scales~\circled{1}.
These shards are retained across microbatches~\circled{2}.

For forward and backward computation, all-gather prepares the
full runtime weights and scales as needed \circled{3}.
The GEMMs consume views of these buffers, and backward writes directly
into the flattened runtime gradient buffer $G'$ \circled{4}.
Reduce-scatter accumulates the gradients in FP32 at ZeRO-2
\circled{5}. This path repeats for each microbatch without repeating
weight quantization or scale relayout. Full weight buffers can be
released after use, while the runtime weight shards and accumulated
gradients remain available.

\paragraph{Update}
After all microbatches, $\mathcal{T}_g$ maps the accumulated gradients
to the update representation at ZeRO-2, and the outer reduce-scatter
produces the ZeRO-1 main-gradient shards \circled{7}.
For Head0, the update spans ranks~2 and~1 \circled{6}.
Each rank first executes $f_0$ on its local fragments to update momentum
and produce BF16 $u$. Rank~2 then gathers the missing fragments of $u$
from rank~1, executes Newton--Schulz $f_1$, and returns the corresponding
FP32 $v$ fragments to rank~1. Both ranks execute $f_2$ locally to update
their main-weight fragments. Main weights and momentum remain sharded
throughout this matrix stage. Applying $\mathcal{G}$ to the updated
main weights begins the next cycle.

The design above provides storage and sharding abstractions for
decoupled runtime and update parameters.
The next section builds on these abstractions to batch updates and
transforms, eliminate redundant copies, and reduce contention between
communication and computation.

%% file: src/4-optimization.tex
\section{Optimization}
\label{sec:optimization}

\begin{figure*}[!t]
  \centering
  \includegraphics[width=\textwidth]{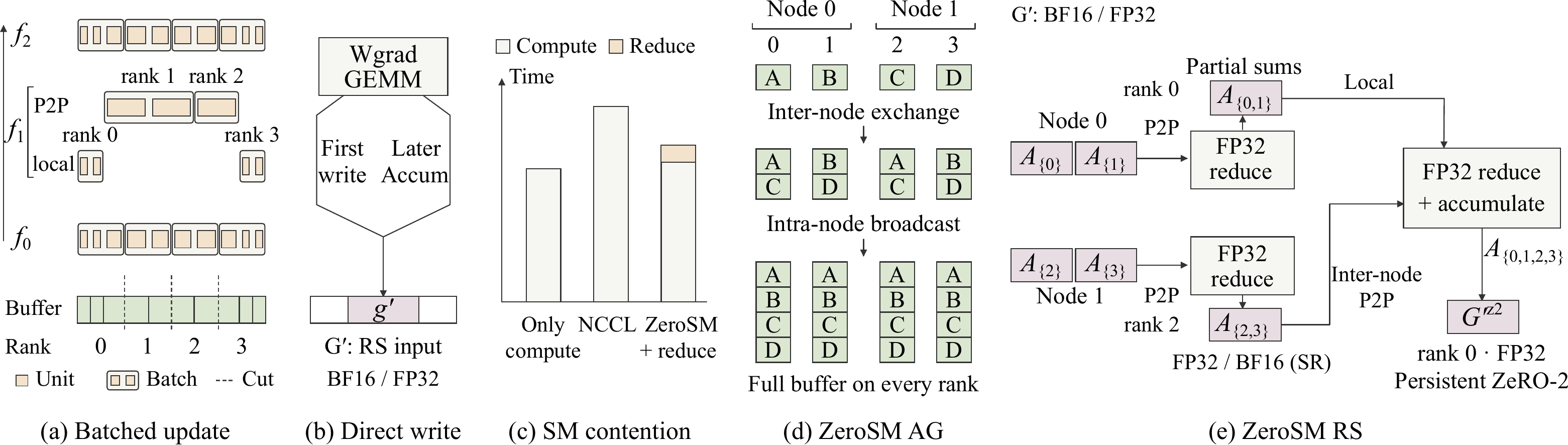}
  \caption{\textbf{Optimizations enabled by runtime--update separation.}
  (a) Compatible update units are batched per stage and execution rank.
  (b) Backward kernels write or accumulate directly into runtime-gradient views.
  (c) SM contention and explicit reduction overhead, shown schematically.
  (d) ZeroSM all-gather exchanges shards between nodes before broadcasting
  within each node.
  (e) Reduce-scatter for one output partition $A$; subscripts identify
  contributing ranks.}
  \label{fig:optimization-overview}
\end{figure*}

The design above enables several optimizations,
summarized in Figure~\ref{fig:optimization-overview}:
batching reduces kernel-launch overhead, shared views and direct gradient
writes avoid redundant memory copies, and ZeroSM collectives reduce
communication-induced SM contention.

\subsection{Batched Update and Transform}
Within each runtime or update group, \our batches compatible operations
of the same type across parameters, as illustrated in
Figure~\ref{fig:optimization-overview}(a).
With independent runtime and update groups,
$\mathcal{T}_w$ and $\mathcal{T}_g$ can be batched within each layer,
while $\mathcal{F}$ and $\mathcal{G}$ can be batched across the model.
Within each stage, compatible local and P2P-dependent units are batched
separately. In our fused Muon implementation, stages $f_0$ and $f_2$
use separate multi-tensor kernels, while $f_1$ groups matrices with
matching shapes and layouts for batched GEMMs. We similarly batch
MXFP scale-layout transforms across tensors. 
Batching reduces kernel-launch overhead across parameters, particularly
for small per-head updates and scale-layout transforms.

\subsection{Fully Zero Copy}
\our arranges ZeRO-1 shards in the flattened buffer to match
the group-rank order of hierarchical collectives, avoiding
buffer reordering.
In Figure~\ref{fig:fhsdp-design}, the four shards have owners
$[0,2,1,3]$ in buffer order. Outer all-gathers within $\{0,2\}$ and
$\{1,3\}$ produce two contiguous halves; runtime all-gathers within
$\{0,1\}$ and $\{2,3\}$ then construct the full buffer directly.
Gradient reduce-scatter follows the same layout in reverse, without reordering.

\paragraph{Zero-copy weights}
When $W'$ is a shared-storage view of $W$, 
using the same shard boundaries and rank ownership allows their local
shards to share storage as well.
In particular, $W'^{\mathrm{z1}}$ and
$W'^{\mathrm{z3}}$ can directly share storage with
$W^{\mathrm{z1}}$ and $W^{\mathrm{z3}}$, respectively.
Individual tensors access these tensor lists through sliced views.
All-gather therefore reads directly from the existing local shard
buffers, and forward and backward kernels consume views of the gathered
buffer. 
No intermediate packing buffer or copy-in/copy-out is needed
for weights. Required scale relayouts are still performed.

\paragraph{Direct gradient write}
The same flattened layout supports direct reduce-scatter, 
but gradient production and accumulation can still introduce additional memory passes.
PyTorch autograd produces per-parameter gradient tensors,
which FSDP2 then copies into a packed reduce-scatter input buffer.

As shown in Figure~\ref{fig:optimization-overview}(b), 
\our supports direct gradient writes from compatible backward kernels, 
such as those for linear layers,
into views of the flattened reduce-scatter input buffer. 
This avoids separate per-parameter gradient tensors and the subsequent packing copies.
The first write must initialize the entire target region, including
zero-contribution entries; subsequent contributions accumulate.
Otherwise, the buffer must be cleared before accumulating contributions.
The remaining addition pass for accumulating reduced gradients across
microbatches is fused into the final stage of reduce-scatter, as described next.

\begin{figure*}[t]
  \centering
  {\def\svgwidth{\textwidth}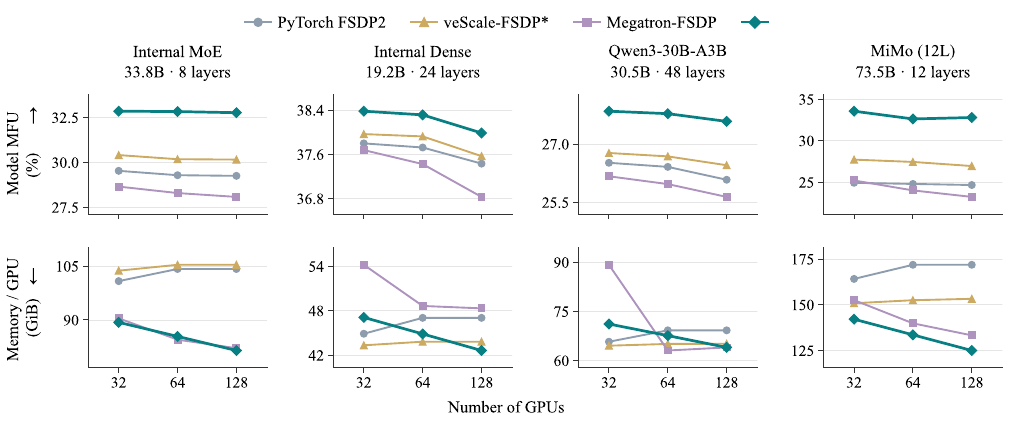}
  \caption{\textbf{MXFP8 training MFU and peak memory at fixed per-GPU workloads.}
  We use $s_w=s_g=32$; for \our and Megatron-FSDP, $s_o$ equals the total GPU count.}
  \label{fig:main-performance-fp8}
\end{figure*}

\subsection{ZeroSM Communication}
\label{sec:zero-sm}

As shown in Figure~\ref{fig:zero-overlap}, communication can overlap with
computation to hide communication latency. In our profiling, NCCL
communication kernels occupy 10--20\% of the SMs.
This creates \emph{SM contention} between communication and computation,
as illustrated in Figure~\ref{fig:optimization-overview}(c).
When ranks progress at different rates, communication kernels may remain
resident while waiting for peer data or synchronization signals, occupying
SM resources even without transferring data.
Table~\ref{tab:kernel-overlap} shows two profiled examples of slower
compute kernels under NCCL overlap.
\begin{table}[t]
  \centering\footnotesize
  \caption{Representative kernel latencies (ms). Parentheses show
  latency increases over standalone execution.}
  \label{tab:kernel-overlap}
  \setlength{\tabcolsep}{3pt}
  \begin{tabular*}{\columnwidth}{@{\extracolsep{\fill}}lrr@{}}
    \toprule
    Kernel & \shortstack[r]{Standalone} & \shortstack[r]{NCCL overlap} \\
    \midrule
    FlashAttention forward & 4.704 & 5.774 (+22.7\%) \\
    DeepGEMM Grouped GEMM & 6.010 & 9.087 (+51.2\%) \\
    \bottomrule
  \end{tabular*}
\end{table}

A fixed SM budget for computation kernels can mitigate contention,
but also limits their parallelism after communication finishes.
We instead use ZeroSM transfers to avoid reserving SM
resources for communication.

\paragraph{ZeroSM P2P}
Within the NVLink domain, GPU copy engines perform peer transfers
without using SMs. Our implementation uses registered symmetric
allocations for communication buffers. Data-tree views address regions
of these registered root buffers, avoiding separate allocations and
registrations for individual views.
Across nodes, we use NCCL Host remote memory access (RMA)~\cite{nccl}.
CPU proxy threads coordinate NIC operations, while RDMA moves the
payload without SM communication kernels.
Together, these mechanisms support intra-node and inter-node ZeroSM P2P
transfers between registered source and destination buffer regions.

\paragraph{All-Gather}
To avoid transferring replicated shards across nodes, we perform
inter-node exchange before intra-node broadcast
(Figure~\ref{fig:optimization-overview}(d)).
Compared with exchanging each GPU's buffer after intra-node broadcast,
this reduces inter-node payload by a factor of $N$, where $N$ is
the number of participating GPUs per node.

\paragraph{Reduce-Scatter}
Unlike all-gather, reduce-scatter still requires arithmetic on SMs.
We use ZeroSM P2P for communication and launch reduction kernels only
after their inputs are ready. These bandwidth-bound kernels can use
the available SMs and HBM bandwidth for local arithmetic without waiting
for remote peers inside the kernel. Figure~\ref{fig:optimization-overview}(c)
illustrates how reduced SM contention can outweigh this explicit reduction cost.

To reduce inter-node traffic, we first aggregate contributions within
each node (Figure~\ref{fig:optimization-overview}(e)). The input buffer
$G'$ holds BF16 or FP32 gradients. Intra-node P2P exchange brings slices
to local reducers, which sum them in FP32, producing partials such as
$A_{\{0,1\}}$ and $A_{\{2,3\}}$.
The partial sums remain FP32 or, for BF16 inputs, are stored in BF16
using stochastic rounding (SR). Inter-node P2P then transfers the partials
to their destination ranks. The final FP32 reduction accumulates directly
into the persistent buffer $G'^{\mathrm{z2}}$, eliminating a separate
result buffer and addition pass.

This follows the separation of BF16 communication and FP32 summation
in DeepSeek-V4~\cite{deepseek-v4}, but applies SR to node-local
partial sums.
For a finite, in-range sum $s$, SR satisfies
$\mathbb{E}[\operatorname{SR}(s)\mid s]=s$:
the rounding error has zero expectation, consistent with unbiased gradient
quantization in low-bitwidth training~\cite{chen2020statistical}.
We write \emph{intra-node/inter-node} for the dtypes of $G'$ and stored node-local
partial sums, the latter also serving as the inter-node payload.
The modes are \textbf{FP32} (FP32/FP32), \textbf{BF16+FP32} (BF16/FP32),
and \textbf{BF16SR} (BF16/BF16).
Section~\ref{sec:precision-ablation} evaluates their effect on training loss.

%% file: figures/main-performance-fp8-v2.pdf_tex
\begingroup%
  \makeatletter%
  \providecommand\color[2][]{%
    \errmessage{(Inkscape) Color is used for the text in Inkscape, but the package 'color.sty' is not loaded}%
    \renewcommand\color[2][]{}%
  }%
  \providecommand\transparent[1]{%
    \errmessage{(Inkscape) Transparency is used (non-zero) for the text in Inkscape, but the package 'transparent.sty' is not loaded}%
    \renewcommand\transparent[1]{}%
  }%
  \providecommand\rotatebox[2]{#2}%
  \newcommand*\fsize{\dimexpr\f@size pt\relax}%
  \newcommand*\lineheight[1]{\fontsize{\fsize}{#1\fsize}\selectfont}%
  \ifx\svgwidth\undefined%
    \setlength{\unitlength}{486bp}%
    \ifx\svgscale\undefined%
      \relax%
    \else%
      \setlength{\unitlength}{\unitlength * \real{\svgscale}}%
    \fi%
  \else%
    \setlength{\unitlength}{\svgwidth}%
  \fi%
  \global\let\svgwidth\undefined%
  \global\let\svgscale\undefined%
  \makeatother%
  \begin{picture}(1,0.42222222)%
    \lineheight{1}%
    \setlength\tabcolsep{0pt}%
    \put(0,0){\includegraphics[width=\unitlength,page=1]{figures/main-performance-fp8-v2.pdf}}%
    \put(0.76460807,0.39483776){\fontfamily{ptm}\fontsize{0.0164609053\unitlength}{0.0205761317\unitlength}\selectfont\color[rgb]{0,0,0}\makebox(0,0)[lt]{\lineheight{1.25}\smash{\begin{tabular}[t]{l}\our{}\end{tabular}}}}%
  \end{picture}%
\endgroup%

%% file: src/5-experiments.tex
\section{Experiments}
\label{sec:experiments}

\subsection{End-to-End Performance}
\label{sec:end-to-end-performance}

\paragraph{Setup}
We compare \our with PyTorch FSDP2,
veScale-FSDP{\renewcommand{\thefootnote}{\fnsymbol{footnote}}\footnote[1]{Results from our reproduction of veScale-FSDP.}},
and Megatron-FSDP on
representative compute-bound training workloads from video and language
models: two internal video models, Qwen3-30B-A3B~\cite{qwen3}, and a
12-layer MiMo~\cite{mimo-v2-flash} variant. We scale from 32 to 128 GPUs with fixed per-GPU
workloads. Detailed settings and measurement procedures are provided in
Appendix~\ref{sec:performance-settings}.

\paragraph{Results}
Figure~\ref{fig:main-performance-fp8} reports the MXFP8 results.
The upper panels show model FLOPs utilization (MFU)~\cite{palm},
defined as achieved model FLOPs per second divided by peak hardware
FLOPs per second. The lower panels show peak memory usage per GPU.

\our achieves the highest MFU in all tested MXFP8 configurations.
At 128 GPUs, the largest relative MFU gain over the fastest baseline
is 21.7\% on MiMo.
MiMo's large expert-weight volume makes weight quantization and copying costly.
Weight reuse avoids repeated quantization across microbatches, while zero-copy
execution eliminates redundant weight and gradient copies.
These optimizations contribute to its larger performance gain.

Although retaining quantized weight shards increases persistent memory usage,
outer ZeRO-1 sharding reduces the memory occupied by main weights and
optimizer states. At 128 GPUs, peak memory is comparable to or lower than
the lowest-memory baseline on all four models.
On MiMo, \our uses 125.0\,GiB per GPU, 6.2\% less than the lowest-memory baseline.

\paragraph{FP8 versus BF16}
MXFP8 improves \our's throughput over BF16 by 16.3--18.5\% on MiMo
and about 7\% on Internal MoE, while Qwen3 throughput changes by less
than 1\% (Appendix~\ref{sec:bf16-performance}).
Qwen3 profiles show that quantization overhead can offset the savings
from faster expert GEMMs. Long-context attention and normalization
times remain the same when expert GEMMs use FP8.

\begin{table}[t]
  \caption{\textbf{Muon versus AdamW on 128 GPUs with BF16 computation.}
  Peak memory (AdamW/Muon, GiB) includes PyTorch reserved memory
  and external communication buffers.}
  \label{tab:muon-overhead}
  \centering\footnotesize
  \setlength{\tabcolsep}{1.5pt}
  \begin{tabular*}{\columnwidth}{@{\extracolsep{\fill}}llrrrr@{}}
    \toprule
    & & \multicolumn{2}{c}{MFU (\%)} & & \\
    \cmidrule(lr){3-4}
    Model & $s_w$ & AdamW & Muon & Overhead & Peak mem. (A/M) \\
    \midrule
    Internal MoE & 128 & 30.34 & 29.60 & +2.5\% & 85.8/82.1 \\
                 & 64  & 30.28 & 29.40 & +3.0\% & 88.8/85.1 \\
                 & 32  & 30.37 & 29.25 & +3.9\% & 92.7/89.0 \\
    \midrule
    Qwen3        & 128 & 27.41 & 26.34 & +4.1\% & 60.1/59.2 \\
                 & 64  & 27.33 & 26.25 & +4.1\% & 62.8/61.9 \\
                 & 32  & 27.14 & 26.23 & +3.5\% & 66.4/65.5 \\
    \bottomrule
  \end{tabular*}
\end{table}

\subsection{Distributed Muon}
\label{sec:distributed-muon}

Using distributed Muon instead of AdamW increases median step time
by only 2.5--4.1\% across the tested models and sharding configurations
(Table~\ref{tab:muon-overhead}). We compare paired runs at 128 GPUs
with the same model, token count, sharding layout.
Muon updates 88.1\% of parameters in Internal MoE and 97.9\% in Qwen3;
AdamW updates the remainder. The implementation keeps main weights and momentum sharded: elementwise stages run locally, while matrix stages exchange only the required intermediate fragments.
 These results show that Muon can be applied to most model parameters with modest end-to-end training overhead.

%% file: src/6-analysis.tex
\section{Ablation and Analysis}
\label{sec:analysis}

\subsection{Precision Ablation}
\label{sec:precision-ablation}

To check that BF16SR in Section~\ref{sec:zero-sm} does not change training convergence,
we ablate communication precision and train a 12-layer, 8.1B-A1.3B
Qwen3 MoE variant~\cite{qwen3} on FineWeb-Edu~\cite{fineweb} for 100B tokens on 128 GPUs with $(s_w,s_g,s_o)=(8,8,128)$.

\begin{figure}[t]
  \centering
  \includegraphics[width=\columnwidth]{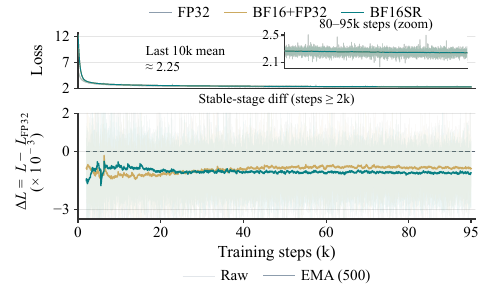}
  \caption{\textbf{Convergence and loss differences.}
  All runs converge without sustained divergence.
  Final-10k-step mean loss ranks \textbf{BF16SR $<$ BF16+FP32 $<$ FP32}.
  Top: training loss. Bottom: $\Delta L=L-L_{\mathrm{FP32}}$.
  Solid lines show EMA (span 500).}
  \label{fig:precision-ablation}
\end{figure}

We compare intra/inter-node reduction input precisions:
\textbf{FP32}, \textbf{BF16+FP32}, and
\textbf{BF16SR}.
All experiments use BF16 runtime weights and FP32 optimizer state.
The three experiments use identical model, data order, optimizer settings, and
seed; only the reduction precision changes.
Figure~\ref{fig:precision-ablation} uses FP32 as the loss-difference
baseline.

All three experiments converge with training-loss differences on the order of $10^{-3}$.
The final-10k-step mean loss ranks
\textbf{BF16SR} $<$ \textbf{BF16+FP32} $<$ \textbf{FP32},
with differences of $-1.10\!\times\!10^{-3}$ and $-8.39\!\times\!10^{-4}$
from FP32, both below 0.05\% in magnitude and stable over training.

\subsection{ZeroSM Communication across Workloads}
\label{sec:zerosm-load}

We vary per-GPU workload size for MiMo at 64 GPUs with
sharding degree 64 under BF16 computation and four microbatches per
optimizer step, comparing ZeroSM, SM-based communication, and default
FSDP2 (Appendix~\ref{sec:zerosm-load-settings}).

\begin{figure}[t]
  \centering
  {\def\svgwidth{\columnwidth}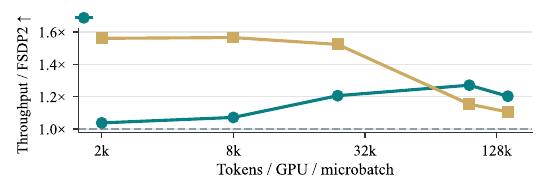}
  \caption{\textbf{The benefit of ZeroSM depends on the workload.}
  Throughput relative to default FSDP2 for MiMo
  (12 layers), using median step times.
  The dashed line marks default FSDP2; the x-axis (tokens per GPU per
  microbatch) is logarithmic.}
  \label{fig:zerosm-load}
\end{figure}

\paragraph{ZeroSM's advantage emerges at larger workloads}
Figure~\ref{fig:zerosm-load} shows a crossover. At the
smallest workload, ZeroSM delivers 33.5\% lower throughput
than SM-based \our{}; at the largest, it gains 8.8\%. The trend is
consistent with a latency-versus-contention trade-off: communication-bound
loads expose collective latency on the critical path, whereas compute-bound
loads hide it through overlap and free SM capacity for model kernels.

\paragraph{SM-based communication also outperforms FSDP2}
SM-based communication retains zero-copy all-gather weight views and
direct-gradient writes. Across the five tested workloads, it achieves
1.106--1.566$\times$ the throughput of default FSDP2, including
1.561$\times$ at the smallest workload. Thus, the throughput advantage
over FSDP2 persists without ZeroSM.

\subsection{Outer ZeRO-1 Sharding and Weight Reuse}
\label{sec:memory-retention}

Outer ZeRO-1 sharding reduces persistent memory by further partitioning FP32
main weights and optimizer states, without enlarging the groups for repeated
weight all-gathers. In the main experiment, \our and Megatron-FSDP fix $s_w=32$
and set $s_o$ to the GPU count, whereas FSDP2 and veScale-FSDP replicate 32-way
optimizer shards. From 32 to 64 to 128 GPUs, \our's peak per-GPU memory falls
by about 3.9\,GiB per doubling on Internal MoE. On MiMo, it falls by about
8.6\,GiB per doubling (Figure~\ref{fig:main-performance-fp8}). Outer ZeRO-1
sharding yields greater memory savings for larger models.
Weight reuse avoids repeated quantization and scale relayout by retaining
low-precision weight shards and scales across microbatches. This requires
additional persistent storage, but at 128 GPUs, \our's peak memory remains
comparable to or lower than the lowest-memory baseline on all tested MXFP8
workloads in Section~\ref{sec:experiments}.

%% file: figures/zerosm-load-v2/a-tradeoff-mimo.pdf_tex
\begingroup%
  \makeatletter%
  \providecommand\color[2][]{%
    \errmessage{(Inkscape) Color is used for the text in Inkscape, but the package 'color.sty' is not loaded}%
    \renewcommand\color[2][]{}%
  }%
  \providecommand\transparent[1]{%
    \errmessage{(Inkscape) Transparency is used (non-zero) for the text in Inkscape, but the package 'transparent.sty' is not loaded}%
    \renewcommand\transparent[1]{}%
  }%
  \providecommand\rotatebox[2]{#2}%
  \newcommand*\fsize{\dimexpr\f@size pt\relax}%
  \newcommand*\lineheight[1]{\fontsize{\fsize}{#1\fsize}\selectfont}%
  \ifx\svgwidth\undefined%
    \setlength{\unitlength}{259.20001221bp}%
    \ifx\svgscale\undefined%
      \relax%
    \else%
      \setlength{\unitlength}{\unitlength * \real{\svgscale}}%
    \fi%
  \else%
    \setlength{\unitlength}{\svgwidth}%
  \fi%
  \global\let\svgwidth\undefined%
  \global\let\svgscale\undefined%
  \makeatother%
  \begin{picture}(1,0.33333332)%
    \lineheight{1}%
    \setlength\tabcolsep{0pt}%
    \put(0,0){\includegraphics[width=\unitlength,page=1]{figures/zerosm-load-v2/a-tradeoff-mimo.pdf}}%
    \put(0.18119695,0.29196361){\fontfamily{ptm}\fontsize{0.0281635802\unitlength}{0.0352044753\unitlength}\selectfont\color[rgb]{0,0,0}\makebox(0,0)[lt]{\lineheight{1.25}\smash{\begin{tabular}[t]{l}\our{} · ZeroSM\end{tabular}}}}%
    \put(0,0){\includegraphics[width=\unitlength,page=2]{figures/zerosm-load-v2/a-tradeoff-mimo.pdf}}%
    \put(0.4862624,0.29196361){\fontfamily{ptm}\fontsize{0.0281635802\unitlength}{0.0352044753\unitlength}\selectfont\color[rgb]{0,0,0}\makebox(0,0)[lt]{\lineheight{1.25}\smash{\begin{tabular}[t]{l}\our{} · SM-based\end{tabular}}}}%
    \put(0,0){\includegraphics[width=\unitlength,page=3]{figures/zerosm-load-v2/a-tradeoff-mimo.pdf}}%
  \end{picture}%
\endgroup%

%% file: src/7-related-work.tex
\section{Related Work}
\label{sec:related-work}

\paragraph{Large-model pretraining practices}
Training language models~\cite{llama3,qwen3,deepseek-v4} and video
models~\cite{vidu,wan,hunyuanvideo-1.5,seedance2} requires flexibility in
sharding, runtime numerical representations, and optimizer update granularity.
DeepSeek-V4 uses different ZeRO partitions for dense and expert parameters
to support Muon, while DeepSeek-V3~\cite{deepseek-v3} and
MiMo-V2-Flash~\cite{mimo-v2-flash} combine FP8 computation with
higher precision for selected operations and parameters.
Kimi K3~\cite{kimi-k3} and GLM-5~\cite{glm5} use attention heads as
Muon update units.

\paragraph{Runtime optimizations for sharded training}
Runtime optimizations target communication volume, memory traffic, and
GPU resource contention during forward and backward execution.
MiCS~\cite{mics} and ZeRO++~\cite{zero-plus-plus} improve communication
locality through smaller sharding groups and intra-node backward weight
all-gathers, respectively.
TorchTitan~\cite{torchtitan} gathers weights in FP8 for subsequent
low-precision computation.
Collective bucketing and reordering~\cite{simplefsdp} and topology-aware
partitioning~\cite{centauri} improve overlap; DeepCompile~\cite{deepcompile}
prefetches parameters to overlap all-gathers with computation and retains
gathered weights to avoid repeated all-gathers.
Computation--collective fusion avoids intermediate memory accesses~\cite{coconet}.
Host-driven NCCLX and NCCL paths offload data movement to reduce
SM usage~\cite{ncclx,nccl-zero-sm}.

\paragraph{Distributed parameter updates}
Distributed optimizer implementations coordinate matrix inputs,
update computation, and persistent states across ranks.
Distributed Muon~\cite{moonlight} gathers matrix inputs for
orthogonalization while retaining sharded optimizer states.
Canzona~\cite{canzona} and DMuon~\cite{dmuon} assign matrix updates to
designated ranks and balance optimizer workloads.
Canzona further pipelines input reconstruction, update computation, and
result redistribution under tensor parallelism.
LinS~\cite{lins} and InternEvo~\cite{internevo} independently configure
parameter, gradient, and optimizer-state sharding to trade memory
for communication.

%% file: src/8-conclusion.tex
\section{Conclusion}
\label{sec:conclusion}

\our shows that flexible optimizer updates and efficient runtime execution can be supported within a unified sharded training framework. Shared storage and hierarchical sharding allow structured updates to coexist with packed computation and communication layouts while avoiding redundant data copies. Our evaluation demonstrates that this flexibility can be achieved with improved training efficiency and competitive memory usage. These findings support separating optimizer update requirements from runtime data organization as a practical approach to sharded training.

%% file: src/a-implementation.tex
\section{Implementation Details}
\label{sec:implementation-details}

\paragraph{Mapping paper terms to the source.}
A logical parameter is represented by \texttt{Parameter}; the kernel-facing
runtime tensor is carried by \texttt{RuntimeParameter}. The source identifiers
\texttt{zero3}, \texttt{zero2}, and \texttt{zero1} denote the retained weight,
accumulation, and optimizer shard levels, respectively. Their paper symbols are
$s_w$, $s_g$, and $s_o$. The copy-engine/RMA implementation resides under
\texttt{zero\_sm}; \texttt{ZeroSmGrad} selects its gradient lifecycle and precision
modes. These identifiers are implementation names rather than additional
parameter or sharding concepts.

\paragraph{Packing and compatibility.}
\texttt{TensorList} stores member shapes and offsets and derives the fragments
at each shard level. An \texttt{Atom} groups parameters with matching tensor
field names, dtypes, devices, optimizer-state layouts, and update buffers.
AdamW and Muon parameters with different state schemas therefore use
separate \texttt{Atom} objects. Runtime and update groups batch compatible operations
across these objects. Each input or output field declares the number of
elements per update unit. Fields may use different numbers of elements
per unit, but must describe the same number of units.

\paragraph{Allocations and update execution.}
The data tree uses \texttt{DataNode} objects whose root owns storage. Finalization
resolves child offsets and forwards allocation callbacks to that root. These
callbacks attach or detach the runtime tensor's view while retaining its
metadata. Update
intermediates such as $u_i$ and $v_i$ have update-scoped allocations; executors retaining
temporary views are released or rebuilt with those allocations. Stages run in
order within one update group, committing intermediates and scalar state for
the next stage. Independent groups retain separate schedules.

\paragraph{Execution conditions.}
Table~\ref{tab:implementation-conditions} distinguishes metadata checks from
execution conditions whose correctness depends on the producer and scheduler.
The direct gradient write mechanism and BF16 communication path are described in
Section~\ref{sec:optimization}; the table records the corresponding storage,
hardware, and restart conditions. In particular, explicit SUM normalization
and valid gradients for every updated logical parameter connect the runtime's
accumulated values to the intended optimizer step.

\begin{table*}[t]
\centering
\caption{Execution conditions, their rationale, and the consequence of violating them. Checks are identified where implemented.}
\label{tab:implementation-conditions}
\label{tab:execution-conditions}
\footnotesize
\renewcommand{\arraystretch}{1.08}
\begin{tabularx}{\textwidth}{@{}>{\raggedright\arraybackslash}p{0.30\textwidth}>{\raggedright\arraybackslash}X>{\raggedright\arraybackslash}X@{}}
\toprule
Condition & Rationale & Consequence if violated \\
\midrule
Direct writes cover every element on the first write of the current buffer lifetime; later writes accumulate.
& Overwrite mode replaces clearing; zero-contribution slices receive explicit zeros.
& Unwritten slices retain stale data, or repeated overwrites discard contributions. \\
Compatible field schemas within each \texttt{Atom}; transforms preserve their declared complete units.
& Packing connects matching fields, dtypes, devices, optimizer states, and update buffers.
& Planning rejects incompatible metadata; incompatible fields use separate \texttt{Atom} objects. \\
Ordered stages within each update group; intermediates are produced before use and scalar updates agree across fragments.
& Each stage consumes committed outputs; independent groups have separate schedules.
& Reads of temporary buffers before production are rejected; inconsistent scalar updates violate stage semantics. \\
Nested sharding degrees; equal shards for copy-engine/RMA collectives; balanced node-major ranks and stable registered bases/offsets.
& Collective layouts and registered addresses describe the same rank-to-data mapping.
& Invalid geometry and detected binding changes are rejected; lifetime violations risk invalid accesses. \\
Events or stream order connect producers, consumers, and reuse; shared RS workspace calls are serialized.
& One arena region serves successive nonoverlapping lifetimes.
& Concurrent reuse races with unfinished transfers or reductions. \\
BF16 partials use supported hardware SR; seed/counter state is restored on the same logical rank.
& The SR instruction and element-to-counter mapping define the numerical path.
& Unsupported hardware or compiled images are rejected before communication; missing RNG state changes resumed rounding. \\
BF16 RS uses full-tensor, layout-preserving Cast transforms, matching packed offsets, and separate BF16 input and FP32 accumulation storage.
& Reduction consumes the input directly while accumulation shards persist.
& Incompatible transform/layout metadata and detected storage overlap are rejected. \\
The specialized copy-engine/RMA path runs without CUDA Graph replay.
& Host protocol counters and Python SR counters have no implemented replay semantics.
& Graph capture/replay is unsupported. \\
All contributing shared-parameter forwards are registered before their final-microbatch backwards.
& Backward-use counts determine eligibility for final synchronization.
& Sequential forward/backward/forward/backward reuse can synchronize too early. \\
Every updated trainable logical parameter has a valid final gradient; new accumulation rounds clear prior gradients explicitly.
& Updates consume gradient shards, without unused-parameter discovery or missing-gradient skipping.
& Missing or stale contributions enter the update. \\
Loss, replica, and microbatch normalization are explicit after final synchronization.
& Gradient collectives use SUM semantics.
& Omitting the intended scale changes the effective gradient. \\
\bottomrule
\end{tabularx}
\end{table*}

%% file: src/b-evaluation.tex
\section{Evaluation and Artifact Details}
\label{sec:artifact}

\begin{figure*}[t]
  \centering
  {\def\svgwidth{\textwidth}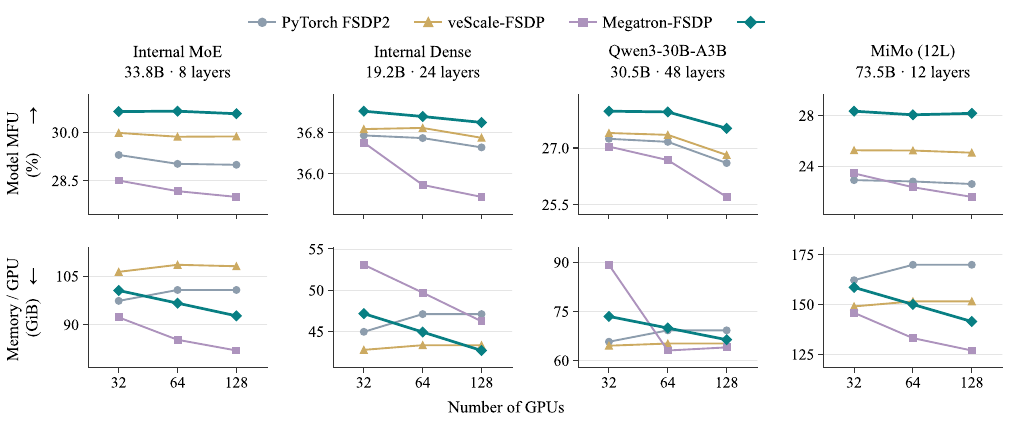}
  \caption{\textbf{BF16 training performance.}
  Workloads and sharding layouts match
  Figure~\ref{fig:main-performance-fp8}, with BF16 GEMMs.
  Top: per-GPU model FLOPs utilization (MFU). Bottom: peak reserved memory
  plus external communication buffers.
  Each model uses its own y-axis ranges.}
  \label{fig:main-performance-bf16}
\end{figure*}

\paragraph{Implementation size.}
The inspected source snapshot contains 85 Python and native source files,
comprising 21,624 physical lines. A source-oriented count excluding blank
lines, comments, and Python docstrings gives approximately 17,606 lines:
14,217 in Python and 3,389 in native code. Preprocessor directives and
structural delimiters are retained. These totals cover shared infrastructure
and specialized backends. The artifact preparation will package the counting
script, file manifest, and immutable source identifier alongside this
breakdown, so the same rule can be applied to extension examples.

\paragraph{Integration and extension measurements.}
We will provide runnable examples showing how to convert a model,
define update parameters, construct runtime parameters, and configure
sharding.
The four-head query example packs $W_Q=[W_0;W_1;W_2;W_3]$, with
$W_i\in\mathbb{R}^{d_h\times d}$, into a BF16 runtime tensor and retains
FP32 main weights for per-head Muon. Its execution settings follow
Table~\ref{tab:execution-conditions}. Each example includes initialization,
gradient clearing, first/last microbatch flags, normalization, and update.
Engineering measurements count added conversion code, optimizer declarations,
backend-specific code, and integration touchpoints for AdamW, expert-wise
Muon, and per-head Muon. Shared utilities are reported separately. Correctness
checks compare the declared logical updates with a local reference.

\paragraph{Reproduction records.}
Machine-readable configurations will record hardware, software versions,
source revisions, topology, rank mapping, model and optimizer settings,
sharding degrees, grouping policy, precision, seeds, and timing boundaries.
Raw per-step times and memory observations accompany aggregate statistics.
Numerical cases preserve the original inputs, communicated representations,
and rounding seeds where feasible, allowing input and partial quantization
to be examined independently. Diagnostics document failed configurations
and the selection of supported replacements.

\paragraph{Artifact preparation.}
Release preparation covers licensing, dependency specifications, runnable
examples, multi-rank correctness tests, benchmark scripts, and backend
configuration guidance. A manifest links each paper experiment to its
configuration and output records. Together, these materials make extension
size, reproduced behavior, and measured performance traceable to the same
implementation snapshot.

\subsection{Performance Settings}
\label{sec:performance-settings}

Table~\ref{tab:performance-workloads} gives the workloads used in
Section~\ref{sec:end-to-end-performance}. All runs use 32, 64, or 128 GPUs
(eight per node), AdamW, full activation recomputation, and gradient
accumulation (GA) of four microbatches per optimizer step.
Inputs are synthetic. Per-GPU microbatch size (MBS) and input length
remain fixed as GPU count increases.

\begin{table}[H]
  \caption{Performance workloads. Parameter counts include all experts.
  MBS denotes samples per GPU per microbatch; GA is the number of
  microbatches per optimizer step.}
  \label{tab:performance-workloads}
  \centering\small
  \setlength{\tabcolsep}{2pt}
  \begin{tabular*}{\columnwidth}{@{\extracolsep{\fill}}lrrrrr@{}}
    \toprule
    Model & Params (B) & Layers & Tokens & MBS & GA \\
    \midrule
    Internal MoE & 33.8 & 8 & 72,186 & 1 & 4 \\
    Internal Dense & 19.2 & 24 & 72,186 & 1 & 4 \\
    Qwen3 & 30.5 & 48 & 24,576 & 4 & 4 \\
    MiMo (12L) & 73.5 & 12 & 24,576 & 4 & 4 \\
    \bottomrule
  \end{tabular*}
\end{table}

Weight and gradient sharding degrees are fixed at 32; \our and
Megatron-FSDP shard optimizer states across all GPUs. MXFP8 uses DeepGEMM
for expert GEMMs, or the video FFN of Internal Dense; remaining GEMMs use
BF16, with FP32 optimizer states. \our and Megatron-FSDP retain quantized weights,
whereas FSDP2 and veScale-FSDP quantize on demand.

\paragraph{Software.}
All end-to-end runs use PyTorch 2.14.0+cu132, CUDA 13.2,
NCCL 2.30.7, and DeepGEMM 2.6.1b2. The PyTorch FSDP2 baseline is the
implementation in PyTorch 2.14; veScale-FSDP runs from our reproduction; Megatron-FSDP is the implementation in Megatron
Core 0.18.2. Baselines use the same model partition, precision settings, and
per-GPU workload as \our.

Timing uses six-step medians after two warmup steps. Model FLOPs exclude
recomputation. We report per-GPU model FLOPs utilization (MFU), i.e.,
achieved model FLOP/s as a fraction of the GPU's BF16 peak FLOP/s. Memory is peak
PyTorch reserved memory plus external communication buffers, maximized over
measured steps and ranks.

\subsection{BF16 Training Performance}
\label{sec:bf16-performance}

Figure~\ref{fig:main-performance-bf16} complements the MXFP8 results in
Section~\ref{sec:end-to-end-performance} with the same workloads, GPU
counts, and measurement procedure. Table~\ref{tab:fp8-bf16-throughput}
compares \our's MXFP8 and BF16 throughput, using ratios of median step
times. The benefit depends on the workload: it is largest on MiMo,
smaller on the internal models, and negligible on Qwen3.

\begin{table}[H]
  \caption{\our MXFP8 throughput relative to its own BF16 runs.}
  \label{tab:fp8-bf16-throughput}
  \centering\small
  \begin{tabular*}{\columnwidth}{@{\extracolsep{\fill}}lrrr@{}}
    \toprule
    Model & 32 GPUs & 64 GPUs & 128 GPUs \\
    \midrule
    Internal MoE & 1.072$\times$ & 1.071$\times$ & 1.072$\times$ \\
    Internal Dense & 1.031$\times$ & 1.032$\times$ & 1.027$\times$ \\
    Qwen3 & 0.995$\times$ & 0.994$\times$ & 1.002$\times$ \\
    MiMo (12L) & 1.185$\times$ & 1.163$\times$ & 1.165$\times$ \\
    \bottomrule
  \end{tabular*}
\end{table}

\input{src/b-zerosm-load}

%% file: figures/main-performance-bf16-v2.pdf_tex
\begingroup%
  \makeatletter%
  \providecommand\color[2][]{%
    \errmessage{(Inkscape) Color is used for the text in Inkscape, but the package 'color.sty' is not loaded}%
    \renewcommand\color[2][]{}%
  }%
  \providecommand\transparent[1]{%
    \errmessage{(Inkscape) Transparency is used (non-zero) for the text in Inkscape, but the package 'transparent.sty' is not loaded}%
    \renewcommand\transparent[1]{}%
  }%
  \providecommand\rotatebox[2]{#2}%
  \newcommand*\fsize{\dimexpr\f@size pt\relax}%
  \newcommand*\lineheight[1]{\fontsize{\fsize}{#1\fsize}\selectfont}%
  \ifx\svgwidth\undefined%
    \setlength{\unitlength}{486bp}%
    \ifx\svgscale\undefined%
      \relax%
    \else%
      \setlength{\unitlength}{\unitlength * \real{\svgscale}}%
    \fi%
  \else%
    \setlength{\unitlength}{\svgwidth}%
  \fi%
  \global\let\svgwidth\undefined%
  \global\let\svgscale\undefined%
  \makeatother%
  \begin{picture}(1,0.42222222)%
    \lineheight{1}%
    \setlength\tabcolsep{0pt}%
    \put(0,0){\includegraphics[width=\unitlength,page=1]{figures/main-performance-bf16-v2.pdf}}%
    \put(0.76049284,0.39483776){\fontfamily{ptm}\fontsize{0.0164609053\unitlength}{0.0205761317\unitlength}\selectfont\color[rgb]{0,0,0}\makebox(0,0)[lt]{\lineheight{1.25}\smash{\begin{tabular}[t]{l}\our{}\end{tabular}}}}%
  \end{picture}%
\endgroup%

%% file: src/b-zerosm-load.tex
\subsection{Communication-Backend Workload Sweep}
\label{sec:zerosm-load-settings}

Section~\ref{sec:zerosm-load} uses a separate BF16 sweep, not the
MXFP8 scaling experiment. MiMo (12 layers, 73.5B parameters) runs on
eight B200 nodes (eight GPUs per node) for each point, with all three
configurations on the same hosts and input shape; otherwise the setup follows
Appendix~\ref{sec:performance-settings}, with DeepGEMM BF16 and
seed~42. Context and expert parallelism are disabled; profiling and
checkpointing are disabled during timing. MiMo uses MBS~1 at 2,048, 8,192, and 24,576,
then MBS~4 at 24,576 and 36,864 (98,304 and 147,456 tokens per GPU per
microbatch). Token count is a workload descriptor, not an
arithmetic-intensity classification.

The sweep has 15 configurations. Each point is the
median of zero-based steps 5--14 (ten timed steps after five warmup
steps), from one run per configuration; these are within-run samples, not
independent repetitions. All accepted runs passed configuration,
host-identity, and GPU-clean checks, with no OOM or allocation retry.

ZeroSM uses BF16 runtime gradients with FP32 inter-node reduction and
accumulation; SM-based communication uses FP32 gradients and NCCL collectives. Both
retain \our's shared weight views and direct gradient writes, but
differ in gradient storage, symmetric arenas, hierarchical reduction,
and accumulation. ZeroSM offloads payload movement while intra-node and
final reductions still use SMs, so the comparison does not isolate
transport latency or SM occupancy. Both \our modes use
$(s_w,s_g,s_o)=(64,64,64)$, and FSDP2 uses sharding degree 64; the layout was
fixed rather than varied, so gains are measured at this sharding degree
only. \our performs
the gradient-to-optimizer reduction only on the final microbatch, and
no deferred-replica-all-reduce variant is evaluated.